# The Ewald–Oseen Extinction Theorem

Masud Mansuripur

College of Optical Sciences, The University of Arizona, Tucson, Arizona 85721

<masud@optics.arizona.edu>

When a beam of light enters a material medium, it sets in motion the resident electrons, whether these electrons are free or bound. The electronic oscillations in turn give rise to electromagnetic radiation which, in the case of linear media, possess the frequency of the exciting beam. Because Maxwell's equations are linear, one expects the total field at any point in space to be the sum of the original (exciting) field and the radiation produced by all the oscillating electrons. However, in practice the original beam appears to be absent within the medium, as though it had been replaced by a different beam, one having a shorter wavelength and propagating in a different direction. The Ewald-Oseen theorem[1,2] resolves this paradox by showing how the oscillating electrons conspire to produce a field that exactly cancels out the original beam everywhere inside the medium. The net field is indeed the sum of the incident beam and the radiated field of the oscillating electrons, but the latter field completely masks the former.[3,4]

Although the proof of the Ewald-Oseen theorem is fairly straightforward, it involves complicated integrations over dipolar fields in three-dimensional space, making it a brute-force drill in calculus and devoid of physical insight.[5,6] It is possible, however, to prove the theorem using plane-waves interacting with thin slabs of material, while invoking no physics beyond Fresnel's reflection coefficients. (These coefficients, which date back to 1823, predate Maxwell's equations.) The thin slabs represent sheets of electric dipoles, and the use of Fresnel's coefficients allows one to derive exact expressions for the electromagnetic field radiated by these dipolar sheets. The integrations involved in this approach are one-dimensional, and the underlying procedures are intuitively appealing to practitioners of optics. The goal of the present article is to outline a general proof of the Ewald-Oseen theorem using arguments that are based primarily on thin-film optics.

**Dielectric slab**. Consider the transparent slab of dielectric material of thickness $d$ and refractive index $n$, shown in Figure 1. A normally incident plane-wave of vacuum wavelength $\lambda_0$ produces a reflected beam of amplitude $r$ and a transmitted beam of amplitude $t$. Both $r$ and $t$ are complex numbers in general, having a magnitude and a phase angle. Using Fresnel's coefficients at each facet of the slab and accounting for multiple reflections, it is fairly straightforward to obtain expressions for $r$ and $t$. The reflection and transmission coefficients at the front facet of the slab are[5,7]

$$\rho = (1 - n)/(1 + n). \tag{1}$$

$$\tau = 2/(1 + n). \tag{2}$$

At the rear facet the corresponding entities are

$$\rho' = (n-1)/(n+1). \tag{3}$$

$$\tau' = 2n/(n+1). \tag{4}$$

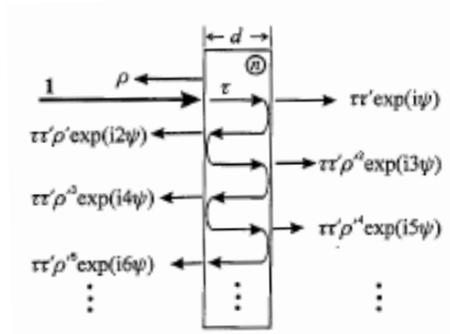

**Figure 1**. Transparent slab of homogeneous material of thickness $d$ and refractive index $n$. A normally-incident monochromatic plane-wave of wavelength $\lambda_0$ suffers multiple reflections at the two facets of the slab. By adding the various reflected and transmitted amplitudes one obtains the expressions for the total $r$ and $t$ given in Eqs.(6) and (7).

A single path of the beam through the slab causes a phase shift $\psi$, where

$$\psi = 2\pi n d/\lambda_0. \tag{5}$$

Adding up all partial reflections at the front facet yields an expression for the reflection coefficient $r$ of the slab. Similarly, adding all partial transmissions at the rear facet yields the transmission coefficient $t$. Thus

$$r = \rho + \tau\tau'\rho'\exp(i2\psi)\sum_{m=0}^{\infty}[\rho'\exp(i\psi)]^{2m} = \rho + \frac{\tau\tau'\rho'\exp(i2\psi)}{1-\rho'^2\exp(i2\psi)}. \tag{6}$$

$$t = \tau\tau'\exp(i\psi)\sum_{m=0}^{\infty}[\rho'\exp(i\psi)]^{2m} = \frac{\tau\tau'\exp(i\psi)}{1-\rho'^2\exp(i2\psi)}. \tag{7}$$

Rather than try to simplify these complicated functions of $n$, $d$ and $\lambda_0$, we show numerical results[8] for the specific case of $n=2$ and $\lambda_0 = 633$nm in Figure 2. The magnitudes of $r$ and $t$ are shown in Figure 2(a), and their phase angles in Figure 2(b), both as functions of the thickness $d$ of the slab. For any given value of $d$ it is possible to represent $r$ and $t$ as complex vectors (see Figure 3). Since the phase difference between $r$ and $t$ is always 90°, these complex vectors are orthogonal to each other. Also, conservation of energy requires that $|r|^2 + |t|^2 = 1$. These observations lead to the conclusion that the hypotenuse of the triangle in Figure 3 must have unit length, that is $|t - r| = 1$, which is also confirmed numerically in Figure 2(c).

Within the slab the incident beam sets the atomic dipoles in motion. These dipoles in turn radiate plane-waves both in forward and backward directions, as shown in Figure 4. When the slab is sufficiently thin, symmetry requires forward- and backward-radiated waves to be identical, that is, they must both have the same amplitude *r*. In the forward direction, however, the incident beam continues to propagate unaltered, except for a phase-shift caused by propagation in free-space through a distance *d*. Thus we must have

$$t = r + \exp(i2\pi d/\lambda_0). \tag{8}$$

It was pointed out earlier in conjunction with the diagram of Figure 3 that $(t - r)$ has unit amplitude, which is in agreement with Eq.(8). It was by no means obvious, however, that the phase of $(t - r)$ must approach $2\pi d/\lambda_0$ as $d \to 0$. Figure 2(c) shows computed plots of the phase of $(t - r)$ normalized by $2\pi d/\lambda_0$. It is seen that in the limit of $d \to 0$ the normalized phase approaches unity as well. This confirms that the slab radiates equally in the forward and backward directions, and that the incident beam, having set the dipolar oscillations in motion, continues to propagate undisturbed in free space.

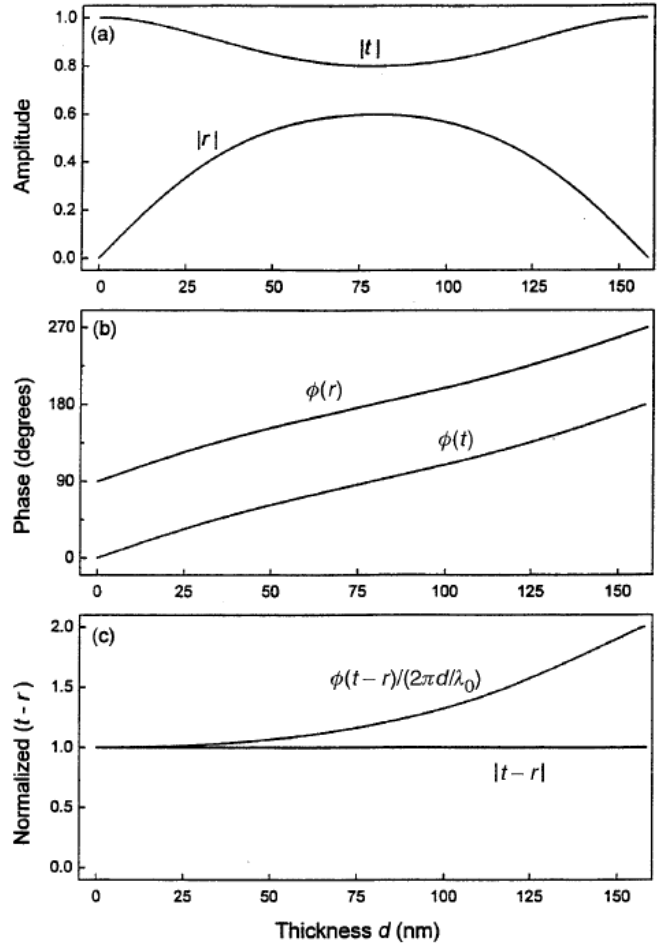

**Figure 2**. Computed plots of *r* and *t* for a slab of thickness *d* and refractive index *n* = 2, when a plane-wave of $\lambda_0$ = 633nm is normally incident on the slab. The horizontal axis covers one cycle of variations of *r* and *t*, corresponding to a half-wave thickness of the slab.

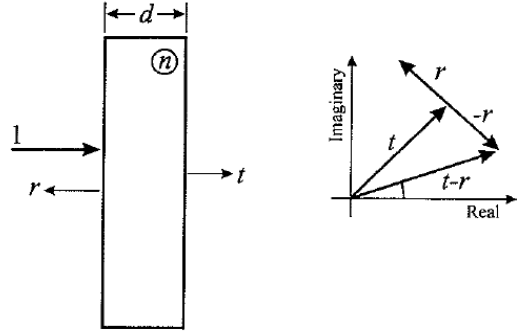

**Figure 3**. Dielectric slab of thickness $d$ and refractive index $n$, reflecting the unit-amplitude incident beam with a coefficient $r$, while transmitting it with a coefficient $t$. The complex-plane diagram at right shows the relative orientations of $r$, $t$ and their difference $(t-r)$. For a non-absorbing slab (i.e., one with a real-valued index $n$) $r$ and $t$ are orthogonal to each other, and $(t-r)$ has unit magnitude.

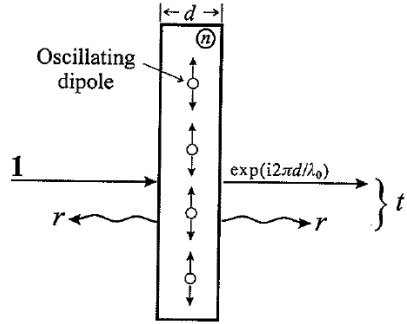

**Figure 4**. Bound electrons within a very thin dielectric slab, when set in motion by a normally incident plane-wave of unit amplitude, radiate with equal strength in both forward and backward directions. The magnitude of the radiated field is the reflection coefficient $r$ of the slab. The incident beam continues to propagate undisturbed as in free-space, acquiring a phase-shift of $2\pi d/\lambda_0$ upon crossing the slab. The sum of the incident beam and the forward-propagating radiated beam constitutes the transmitted beam.

**Radiation from uniform sheet of oscillating dipoles**. In the limit of small $d$, Eq.(6) reduces to the following simple form:

$$r \approx i[\pi(n^2-1)d/\lambda_0]\exp[i\pi(n^2+1)d/\lambda_0]; \qquad (d/\lambda_0) \ll 1. \qquad (9)$$

In this limit the radiated field is slightly more than 90° ahead of the incident field, while its amplitude is proportional to $d/\lambda_0$ and also proportional to $(n^2-1)$, the latter being the coefficient of polarizability of the dielectric material. Note that the small phase angle of $r$ above and beyond its 90° phase, i.e., the exponential factor in Eq.(9), is essential for conservation of energy among the incident, reflected, and transmitted beams (see Figure 4).

Equation (9) is in fact the exact solution of Maxwell's equations for the radiation field of a sheet of dipole oscillators. Although derived here as an aid in proving the extinction theorem, it is an important result in its own right. Note, for example, that the amplitude of the radiated field is proportional to $1/\lambda_0$, even though the field of individual dipole radiators is known to be proportional to $1/\lambda_0^2$. The coherent addition of amplitudes over the sheet of dipoles has thus modified the wavelength-dependence of the radiated field.[4]

**The Extinction Theorem**. Having derived Eq.(9) for the field radiated by a sheet of dipoles, we are now in a position to outline the proof of the extinction theorem. Consider a semi-infinite, homogeneous medium of refractive index $n$, bordering with free-space at $z = 0$, as shown in Figure 5. A unit magnitude plane-wave of wavelength $\lambda_0$ is directed at this medium at normal incidence from the left side. To determine the reflected amplitude $\rho$ at the interface, divide the medium into thin slabs of thickness $\Delta z$, then add up (coherently) the reflected fields from each of these slabs. Similarly, the field at an arbitrary plane $z = z_0$ inside the medium may be computed by adding to the incident beam the contributions of the slabs located to the left of $z_0$ as well as those to the right of $z_0$. The simplest way to proceed is by assuming that the field inside the medium has the expected form, $\tau \exp(\mathrm{i} 2\pi n d/\lambda_0)$, then showing self-consistency. These calculations involve simple one-dimensional integrals, and are in fact so straightforward that there is no need to carry them out here. The interested reader may spend a few minutes to evaluate the integrals and convince himself of the validity of the theorem.

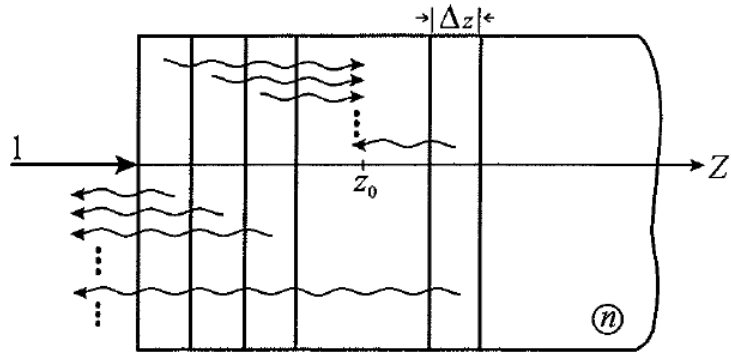

**Figure 5**. A semi-infinite medium of refractive index $n$ is illuminated by a unit-amplitude plane-wave at normal incidence. The medium may be considered as a contiguous sequence of thin slabs, each radiating with equal strength in both forward and backward directions. Adding the backward-radiated fields coherently yields the reflection coefficient at the front facet of the medium. Similarly, the internal field at $z = z_0$ is obtained by coherent addition of the incident beam, the forward-propagating radiations from the left side of $z_0$, and the backward-propagating radiations from the right side of $z_0$.

**Slab of absorbing material**. When the material of the slab happens to be absorbing, similar arguments as above may be advanced to prove the Ewald-Oseen theorem, although the expressions for reflection and transmission coefficients become more complicated. Numerically, however, it is still possible to describe the situation with great accuracy.

Figure 6 shows computed plots[8] of $r$ and $t$ for a metal slab having the complex index $n+\mathrm{i}k=2+\mathrm{i}7$. (Compare these plots with the corresponding plots of the dielectric slab in Figure 2.) It is seen in Figure 6(a) that the reflectance drops sharply while the transmittance increases as the film thickness is reduced below about 20 nm. The phase plots in Figure 6(b) are quite different from those of the dielectric slab, indicating a phase difference greater than 90° between $r$ and $t$. A complex-plane diagram for this type of material is given in Figure 7. The angle between $r$ and $t$ being greater than 90° implies that $|t-r|^2 > |t|^2 + |r|^2$, while conservation of energy requires $|t|^2+|r|^2 < 1$ in the case of absorbing media. The fact that $|t-r|$ can approach unity is borne out by the numerical results depicted in Figure 6(c). In the limit of $d \to 0$, not only does the magnitude of $(t-r)$ become unity, but also its phase approaches $2\pi d/\lambda_0$. Therefore, in the limit of small $d$, the transmitted beam may be expressed as the sum of reflected and phase-shifted incident beams, the phase shift being due to free-space propagation over the distance $d$. This is all that one needs in order to prove the extinction theorem for absorbing media.

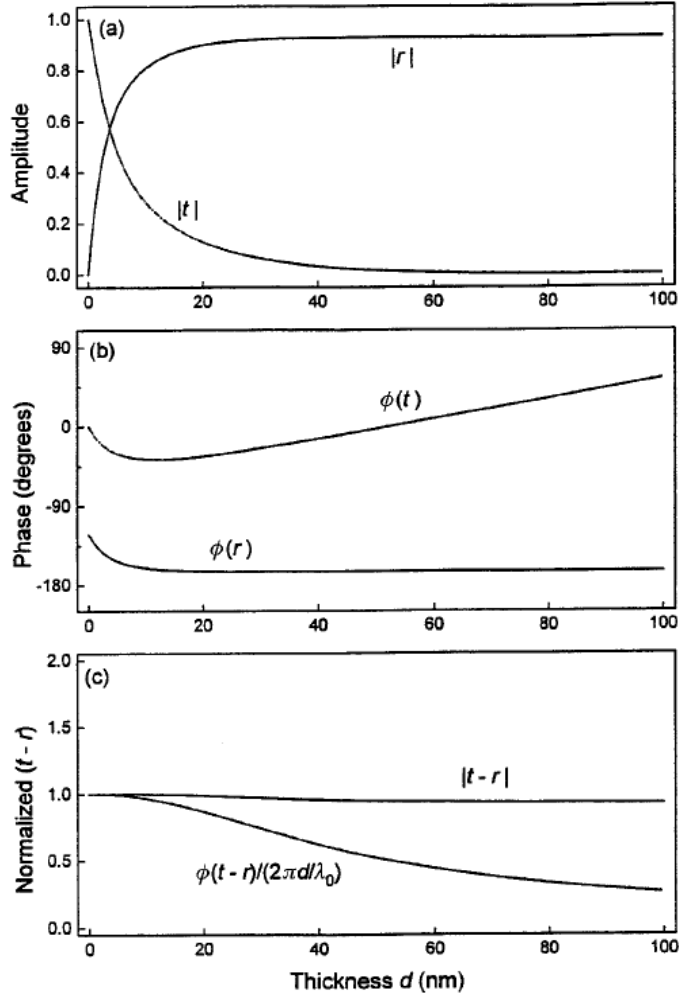

**Figure 6**. Computed plots of $r$ and $t$ for a slab of thickness $d$ and complex refractive index $(n, k) = (2, 7)$, when a plane-wave of $\lambda_0 = 633$ nm is normally incident on the slab. The horizontal axis covers the penetration depth of the material.

**Oblique incidence on a dielectric slab**. Figure 8 shows an s-polarized plane-wave at oblique incidence on a dielectric slab of thickness $d$ and index $n$. The oscillating dipoles are parallel to the s direction of polarization, and radiate with equal magnitude in forward and backward directions. The computed plots of $r_s$ and $t_s$ versus $d$ for the specific case of $\lambda_0 = 633$ nm, $n = 2$, and $\theta = 50°$ are shown in Figure 9. The angle of propagation inside the medium is obtained from Snell's law as $\theta' = 22.52°$, and the half-wave thickness of the slab is given by $\lambda_0/(2n \cos \theta') = 171.3$ nm. These curves are very similar to those of Figure 2, showing a 90° phase difference between $r_s$ and $t_s$, unit magnitude for $(t_s - r_s)$, and a phase for $(t_s - r_s)$ that approaches $2\pi(d/\lambda_0) \cos \theta$ as $d \to 0$. The Ewald-Oseen theorem for the case of s-polarized light at oblique incidence can therefore be proven along the same lines as described earlier for normal incidence.

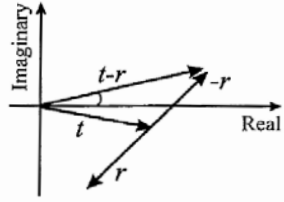

**Figure 7**. Complex-plane diagram showing the reflection coefficient $r$, transmission coefficient $t$, and their difference $(t-r)$ for a thin slab of an absorbing material.

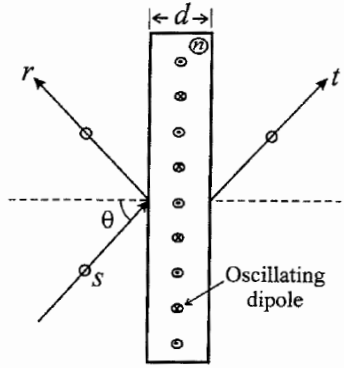

**Figure 8**. An s-polarized plane-wave is obliquely incident at an angle $\theta$ on a dielectric slab of thickness $d$ and index $n$. The electric dipoles of the slab oscillate perpendicular to the plane of the diagram, radiating identical fields in the forward and backward directions.

The case of p-polarized light depicted in Figure 10 is somewhat different, however. Here the directionality of the oscillating dipoles within the slab breaks the symmetry between forward- and backward-radiated beams. The angle $\theta''$ between the direction of oscillating dipoles and the plane of the slab may be determined by considering multiple reflections within the slab. For very thin slabs, it is possible to show that

$$\tan\theta'' = (1/n^2)\tan\theta. \tag{10}$$

Note that at Brewster's angle, where $\tan\theta = n$, we have $\tan\theta'' = 1/n$, that is, $\theta'' = \theta'$, where $\theta'$ is the propagation angle within the medium as given by Snell's law. At angles below the Brewster angle $\theta'' < \theta'$, while above the Brewster angle $\theta'' > \theta'$.

For the case of p-polarized light of wavelength $\lambda_0 = 633$nm incident at $\theta = 50°$ on a slab of index $n = 2$, plots of $r$ and $t$ versus the slab thickness $d$ are shown in Figure 11. Although the magnitude of $(t_p - r_p)$ can still be shown to be unity, its phase does not approach $2\pi(d/\lambda_0)\cos\theta$ as $d \to 0$. This is a manifestation of the breakdown of symmetry between forward and backward radiations. If the relative magnitude of the radiated beams in the two directions is taken into account, however, the preceding arguments can be restored. One may readily observe in Figure 10 that the ratio of forward- to backward-propagating beams must be given by

$$W(\theta) = \cos(\theta - \theta'')/\cos(\theta + \theta''). \tag{11}$$

Therefore, for p-polarized light at oblique incidence, it is $(t - Wr)$ that approaches $\exp(i2\pi d\cos\theta/\lambda_0)$ as $d \to 0$. This is indeed verified in Figure 11(c).

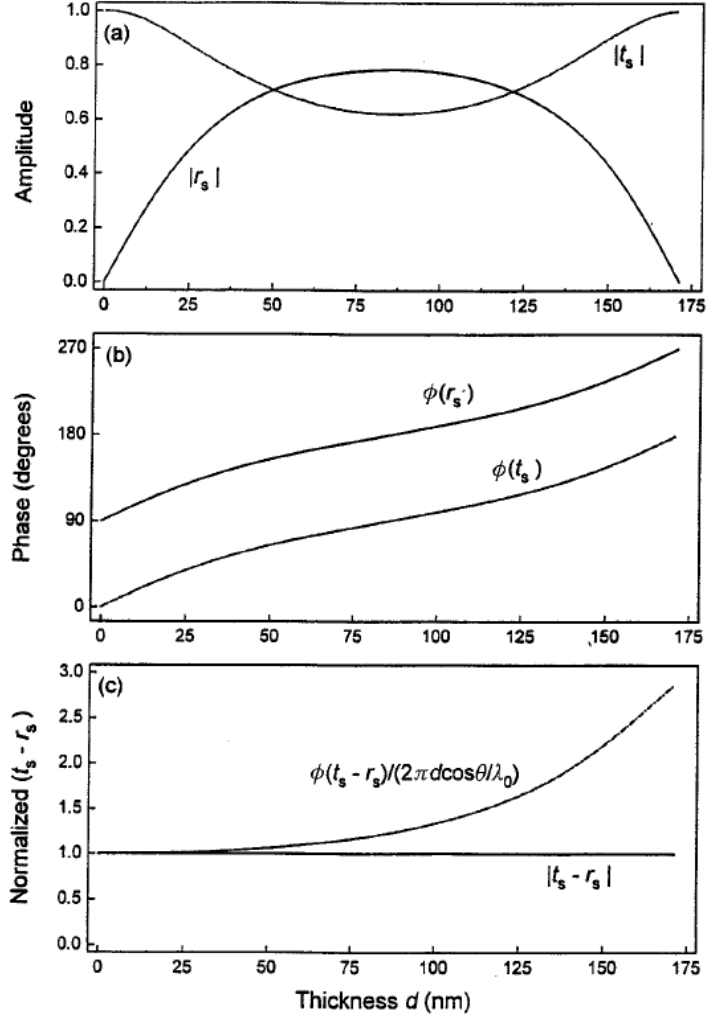

**Figure 9.** Computed plots of $r$ and $t$ for a slab of thickness $d$ and index $n = 2$, when an s-polarized plane-wave of $\lambda_0 = 633$ nm illuminates the slab at the oblique angle of $\theta = 50°$. The horizontal axis covers one cycle of variations of $r$ and $t$, corresponding to a half-wave thickness of the slab at this particular angle of incidence.

As a further test of Eq.(11), we show in Figure 12 the computed plot versus $\theta$ of $r_p/[t_p - \exp(\mathrm{i}2\pi d\cos\theta/\lambda_0)]$ for a slab of $d = 10$ nm and $n = 2$, illuminated with a plane-wave of $\lambda_0 = 633$ nm. This curve overlaps the plot of the function $1/W(\theta)$ exactly. Taking into account the ratio $W(\theta)$ between the forward and backward radiated beams, it is possible once again to prove the Ewald-Oseen theorem as before.[†]

---

[†] See the addendum starting on page 13 for a detailed derivation of the theorem in the case of p-polarized light.

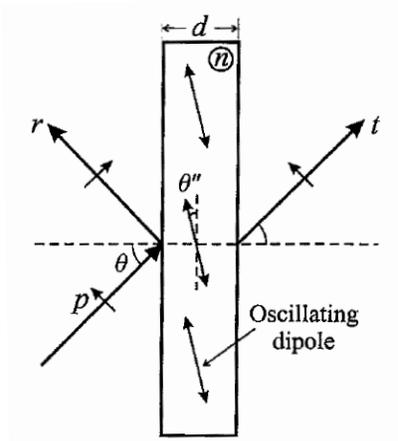

**Figure 10**. A p-polarized plane-wave is obliquely incident at an angle $\theta$ on a dielectric slab of thickness $d$ and index $n$. The oscillating dipoles make an angle $\theta''$ with the surface of the slab, radiating with different amplitudes in the forward and backward directions.

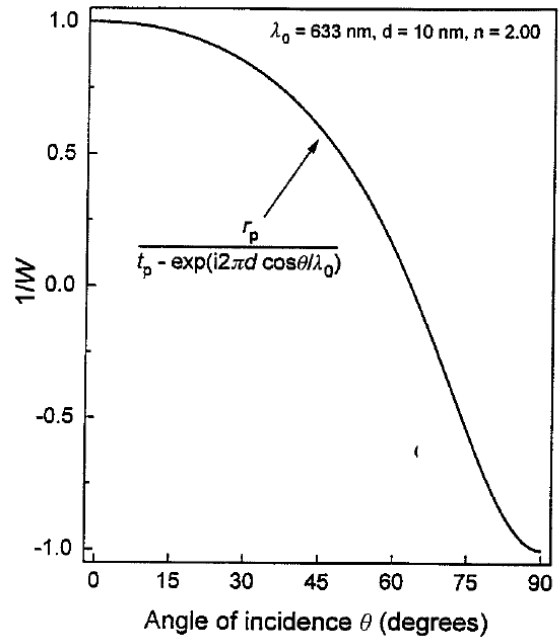

**Figure 12**. Computed ratio of the amplitudes of backward-propagating radiation to forward-propagating radiation for a 10nm-thick dielectric slab of $n = 2$. A p-polarized plane-wave of $\lambda_0 = 633$ nm is assumed to be obliquely incident on the slab at an angle $\theta$.

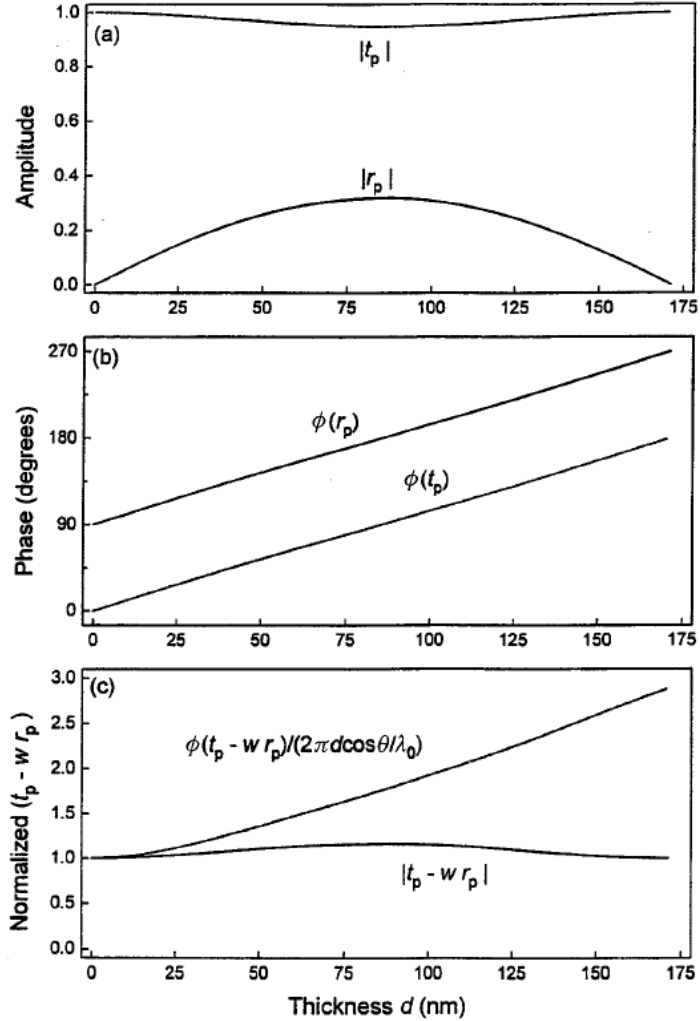

**Figure 11**. Computed plots of *r* and *t* for a slab of thickness *d* and index *n* = 2, when a p-polarized plane-wave of $\lambda_0 = 633$ nm illuminates the slab at the oblique angle of $\theta = 50°$. The horizontal axis covers one cycle of variations of *r* and *t*, corresponding to a half-wave thickness of the slab at this particular angle of incidence.

**This paper, when originally published in *Optics & Photonics News*, prompted the following criticism and reply.**

Editor:

While we are pleased that Masud Mansuripur has called attention in OPN to the rather basic Ewald-Oseen extinction theorem, we wish to take issue with certain parts of his article.[1]

Mansuripur states that the goal of his article is "to outline a general proof of the Ewald-Oseen theorem using arguments that are based primarily on thin-film optics." We wish to note first that the proof he outlines, based on the field produced by a uniform sheet of dipole oscillators and the assumed form $\exp[2\pi i n z/\lambda_0]$ for the field inside the medium, is essentially the same approach used by Fearn, James, and Milonni.[2] Their proof is more general in that Fresnel coefficients (for normal incidence) are derived rather than assumed. Indeed, the derivation of the Fresnel coefficients assumes the extinction of the incident field inside the dielectric medium: Mansuripur's starting point implicitly assumes the very theorem he is trying to prove! In this connection we note that it was not claimed by Fearn *et al.* that they provided a "general proof" of the extinction theorem. A general proof, valid for media bounded by surfaces of arbitrary shape, is given by Born and Wolf.[3]

Mansuripur cites References 2 and 3 in support of his opinion that the proof of the extinction theorem is "devoid of physical insight." While it is true that the proofs given in these references involve "complicated integration over dipolar fields in three-dimensional space," we do not think it is fair to say it is devoid of physical insight. In Reference 3, page 101, the significance of the theorem is described in the following manner that could hardly be more physical: "The incident wave may… be regarded as extinguished at any point within the medium by interference with the dipole field and replaced by another wave with a different velocity (and generally also a different direction) of propagation."

Finally we note that various features of the extinction theorem have been interpreted differently by various authors: some of these differences have been discussed by Fearn *et al.*[2] It would be unfortunate if readers of Mansuripur's article were left with the impression that the theorem can somehow be based "primarily on thin-film optics."

1. M. Mansuripur, "The Ewald-Oseen extinction theorem," *Opt. & Phot. News* **9** (8), 50-55 (1998).
2. H. Fearn *et al.*, "Microscopic approach to reflection, transmission, and the Ewald-Oseen extinction theorem," Am. J. Phy. **64**, 986-995 (1996).
3. M. Born and E. Wolf, *Principles of Optics*, sixth edition (Cambridge University Press, Cambridge, U.K. 1985), section 2.4.2.

Daniel James and Peter W. Milonni, Los Alamos National Laboratory, Los Alamos, NM
Heidi Fearn, California State University at Fullerton, Fullerton, CA
Emil Wolf, University of Rochester, Rochester, NY

**The author replies**:

It is puzzling that Fearn *et al* consider starting from Fresnel's reflection coefficients a shortcoming of my method of proof. The Fresnel coefficients can be derived directly from Maxwell's equations without invoking the extinction theorem, they are available in many textbooks (including Born and Wolf, 6th edition, pp 38-41), and their derivation from first principles does not in any way add to the value of a paper. I used Fresnel's coefficients to derive the radiation field for a sheet of dipoles (Equation 9 of my article), as this is a simple, accurate, and intuitive way of calculating the field, and also because its underlying principle is familiar to many practitioners of optics. Alternatively, one could derive the radiation field by integrating over individual dipoles within the sheet, as is done, for example, in *The Feynman Lectures on Physics* (my reference 3). After this step that establishes the radiation field from a dipolar sheet, the method of proof that I proposed (based on demonstrating self-consistency) is similar to that of Fearn *et al*.

Although Fresnel's coefficients are derived from Maxwell's equations, nowhere in the standard derivation is it assumed that the incident beam is still present within the medium (albeit masked by the dipole radiations). Had the Ewald-Oseen theorem been somehow implicit in the standard derivation of Fresnel's coefficients, there would have been no need for the paper of Fearn *et al* in the first place.

I strongly disagree with the suggestion that the use of Fresnel's coefficients somehow renders my proof of the Ewald-Oseen theorem circular. I also dispute the assertion made by Fearn *et al* that "it would be unfortunate if readers ... were left with the impression that the theorem can somehow be based primarily on thin film optics." Emphatically, the proof of the theorem *can* be based on thin film optics (this is exactly what I showed in the article), and it is far from "unfortunate" indeed when a valid proof happens to be based on a simple physical picture.

I erred in stating that I was going to "outline a general proof of the ... theorem." Mine was a general proof for the one-dimensional case, where the beam enters from free space through a plane boundary into an isotropic, homogeneous medium. My proof is more general than the proof of Fearn *et al*, in that it covers both transparent and absorbing media, and also in that it considers the case of oblique incidence with *p* and *s* polarized light. The method described in Born and Wolf is obviously more general than both, because it applies to arbitrary boundaries. None of the above methods, however, is sufficiently general to embrace inhomogeneous, anisotropic, and optically active media, for which the theorem is presumably valid as well.

Finally, my expressed opinion regarding the proof of the extinction theorem being "devoid of physical insight" was meant as a commentary on the nature of the method, not as a reflection on the authors of the cited references. Ultimately, of course, such judgments are subjective and are best left to the readers.

# Addendum to the Ewald-Oseen Extinction Theorem

Masud Mansuripur

College of Optical Sciences, The University of Arizona, Tucson, Arizona 85721
masud@optics.arizona.edu

In the case of *p*-polarized light, the proof of the Ewald-Oseen extinction theorem requires the inclusion of certain weight factors to account for the differences between the *E* field profile in the thin-film slab and that in a semi-infinite medium. Here we derive all the necessary equations and give a detailed proof of the theorem.

Let the *p*-polarized plane-wave, incident at an angle $\theta$ in the $xz$-plane, have unit amplitude $E_{po} = 1.0$ and vacuum wavelength $\lambda_0$. The *E* and *H* fields inside and outside a slab of thickness *d* and refractive index *n* may be written as follows:

$$\boldsymbol{E}^{(i)}(\boldsymbol{r},t) = (\cos\theta\,\hat{\boldsymbol{x}} - \sin\theta\,\hat{\boldsymbol{z}})\exp[\mathrm{i}(2\pi/\lambda_0)(x\sin\theta + z\cos\theta - ct)], \tag{1a}$$

$$\boldsymbol{H}^{(i)}(\boldsymbol{r},t) = Z_o^{-1}\hat{\boldsymbol{y}}\exp[\mathrm{i}(2\pi/\lambda_0)(x\sin\theta + z\cos\theta - ct)]. \tag{1b}$$

$$\boldsymbol{E}^{(r)}(\boldsymbol{r},t) = r(\cos\theta\,\hat{\boldsymbol{x}} + \sin\theta\,\hat{\boldsymbol{z}})\exp[\mathrm{i}(2\pi/\lambda_0)(x\sin\theta - z\cos\theta - ct)], \tag{2a}$$

$$\boldsymbol{H}^{(r)}(\boldsymbol{r},t) = -Z_o^{-1}r\hat{\boldsymbol{y}}\exp[\mathrm{i}(2\pi/\lambda_0)(x\sin\theta - z\cos\theta - ct)]. \tag{2b}$$

$$\boldsymbol{E}^{(A)}(\boldsymbol{r},t) = E_A(\cos\theta'\,\hat{\boldsymbol{x}} - \sin\theta'\,\hat{\boldsymbol{z}})\exp[\mathrm{i}(2\pi/\lambda_0)(x\sin\theta + nz\cos\theta' - ct)], \tag{3a}$$

$$\boldsymbol{H}^{(A)}(\boldsymbol{r},t) = Z_o^{-1}nE_A\hat{\boldsymbol{y}}\exp[\mathrm{i}(2\pi/\lambda_0)(x\sin\theta + nz\cos\theta' - ct)]. \tag{3b}$$

$$\boldsymbol{E}^{(B)}(\boldsymbol{r},t) = E_B(\cos\theta'\,\hat{\boldsymbol{x}} + \sin\theta'\,\hat{\boldsymbol{z}})\exp[\mathrm{i}(2\pi/\lambda_0)(x\sin\theta - nz\cos\theta' - ct)], \tag{4a}$$

$$\boldsymbol{H}^{(B)}(\boldsymbol{r},t) = -Z_o^{-1}nE_B\hat{\boldsymbol{y}}\exp[\mathrm{i}(2\pi/\lambda_0)(x\sin\theta - nz\cos\theta' - ct)]. \tag{4b}$$

$$\boldsymbol{E}^{(t)}(\boldsymbol{r},t) = t(\cos\theta\,\hat{\boldsymbol{x}} - \sin\theta\,\hat{\boldsymbol{z}})\exp[\mathrm{i}(2\pi/\lambda_0)(x\sin\theta + z\cos\theta - ct)], \tag{5a}$$

$$\boldsymbol{H}^{(t)}(\boldsymbol{r},t) = Z_o^{-1}t\hat{\boldsymbol{y}}\exp[\mathrm{i}(2\pi/\lambda_0)(x\sin\theta + z\cos\theta - ct)]. \tag{5b}$$

The superscripts (i), (r) and (t) in the above equations stand for the incident, reflected, and transmitted beams, whereas the superscripts (A) and (B) indicate the forward and backward propagating plane-waves inside the slab. $Z_o = \sqrt{\mu_o/\varepsilon_o}$ is the impedance of free space, and the refraction angle $\theta'$ is given by Snell's law, $n\sin\theta' = \sin\theta$. Continuity of the tangential *E* and *H* fields at the front and rear facets of the slab yields the following boundary conditions:

$$(1+r)\cos\theta = (E_A + E_B)\cos\theta', \tag{6a}$$

$$(1-r) = n(E_A - E_B), \tag{6b}$$

$$\left[E_A\exp\left(\frac{\mathrm{i}2\pi nd\cos\theta'}{\lambda_0}\right) + E_B\exp\left(-\frac{\mathrm{i}2\pi nd\cos\theta'}{\lambda_0}\right)\right]\cos\theta' = t\cos\theta\exp\left(\frac{\mathrm{i}2\pi d\cos\theta}{\lambda_0}\right), \tag{6c}$$

$$n\left[E_A\exp\left(\frac{\mathrm{i}2\pi nd\cos\theta'}{\lambda_0}\right) - E_B\exp\left(-\frac{\mathrm{i}2\pi nd\cos\theta'}{\lambda_0}\right)\right] = t\exp\left(\frac{\mathrm{i}2\pi d\cos\theta}{\lambda_0}\right). \tag{6d}$$

Before proceeding further, let us specialize the above results to the case of a semi-infinite slab, for which we can set $E_B = 0$ and, consequently, express the front facet Fresnel reflection and transmission coefficients, $\rho = r$ and $\tau = E_A$, as follows:

$$\rho = \frac{\cos\theta' - n\cos\theta}{\cos\theta' + n\cos\theta}, \tag{7a}$$

$$\tau = \frac{2\cos\theta}{\cos\theta' + n\cos\theta}. \tag{7b}$$

Returning now to the case of a finite-thickness slab, we eliminate the transmission coefficient $t$ between Eqs.(6c) and (6d) to find

$$E_B = E_A \left(\frac{n\cos\theta - \cos\theta'}{n\cos\theta + \cos\theta'}\right) \exp\left(\frac{i4\pi nd \cos\theta'}{\lambda_0}\right). \tag{8}$$

Substitution for $E_B$ from Eq.(8) into Eqs.(6a) and (6b) enables one to determine the reflection coefficient $r$, as follows:

$$r = \frac{(n^2\cos^2\theta - \cos^2\theta')[\exp(i4\pi nd\cos\theta'/\lambda_0) - 1]}{(n\cos\theta + \cos\theta')^2 - (n\cos\theta - \cos\theta')^2 \exp(i4\pi nd\cos\theta'/\lambda_0)}. \tag{9}$$

One may now approximate the above $r$ in the limit of a thin slab, $d \ll \lambda_0$, up to and including second order terms in $d/\lambda_0$, as follows:

$$r \cong \frac{(n^2\cos^2\theta - \cos^2\theta')\left[\left(\frac{i4\pi nd\cos\theta'}{\lambda_0}\right) + \tfrac{1}{2}\left(\frac{i4\pi nd\cos\theta'}{\lambda_0}\right)^2\right]}{(n\cos\theta + \cos\theta')^2 - (n\cos\theta - \cos\theta')^2\left[1 + \left(\frac{i4\pi nd\cos\theta'}{\lambda_0}\right) + \tfrac{1}{2}\left(\frac{i4\pi nd\cos\theta'}{\lambda_0}\right)^2\right]}$$

$$= \frac{\left(\frac{i4\pi nd\cos\theta'}{\lambda_0}\right)(n^2\cos^2\theta - \cos^2\theta')\left(1 + \frac{i2\pi nd\cos\theta'}{\lambda_0}\right)}{4n\cos\theta\cos\theta'\left[1 - \frac{(n\cos\theta - \cos\theta')^2}{4n\cos\theta\cos\theta'}\left(\frac{i4\pi nd\cos\theta'}{\lambda_0}\right)\left(1 + \frac{i2\pi nd\cos\theta'}{\lambda_0}\right)\right]}$$

$$\cong \left(\frac{i\pi d}{\lambda_0}\right)\left(\frac{n^2\cos^2\theta - \cos^2\theta'}{\cos\theta}\right)\left(1 + \frac{i2\pi nd\cos\theta'}{\lambda_0}\right)\left[1 + \left(\frac{i\pi d}{\lambda_0}\right)\frac{(n\cos\theta - \cos\theta')^2}{\cos\theta}\left(1 + \frac{i2\pi nd\cos\theta'}{\lambda_0}\right)\right]$$

$$\cong \frac{i\pi d(n^2\cos^2\theta - \cos^2\theta')}{\lambda_0 \cos\theta}\left[1 + \frac{i2\pi nd\cos\theta'}{\lambda_0} + \frac{i\pi d(n\cos\theta - \cos\theta')^2}{\lambda_0 \cos\theta}\right]$$

$$\cong \frac{i\pi d(n^2\cos^2\theta - \cos^2\theta')}{\lambda_0 \cos\theta} \exp\left[\frac{i\pi d(n^2\cos^2\theta + \cos^2\theta')}{\lambda_0 \cos\theta}\right]. \tag{10}$$

In what follows, we will *not* need the exponential factor in the above expression of $r$. We have retained this term for the sake of completeness, as it would be useful in calculations that require the second order term in $d/\lambda_0$. The following analysis, however, does not need this exponential phase-factor.

Inside the slab, the $E$ field is the sum of $\boldsymbol{E}^{(A)}$ and $\boldsymbol{E}^{(B)}$. For an exceedingly thin slab, we may ignore the field's $z$-dependence and assume that $\boldsymbol{E}$ is uniform along the $z$-axis. We find

$$\boldsymbol{E}^{(\text{inside})}(\boldsymbol{r}, t) \cong [(E_A + E_B)\cos\theta'\,\hat{\boldsymbol{x}} - (E_A - E_B)\sin\theta'\,\hat{\boldsymbol{z}}]\exp[i(2\pi/\lambda_0)(x\sin\theta - ct)]$$

$$= [(1 + r)\cos\theta\,\hat{\boldsymbol{x}} - n^{-2}(1 - r)\sin\theta\,\hat{\boldsymbol{z}}]\exp[i(2\pi/\lambda_0)(x\sin\theta - ct)]$$

Dropping $r$ in the limit $d \ll \lambda_0$ → $\cong \cos\theta\,(\hat{\boldsymbol{x}} - n^{-2}\tan\theta\,\hat{\boldsymbol{z}})\exp[i(2\pi/\lambda_0)(x\sin\theta - ct)]. \tag{11}$

Defining the angle $\theta''$ between the internal $E$ field and the $x$-axis by $\tan\theta'' = n^{-2}\tan\theta$, the above equation may now be rewritten as

$$\boldsymbol{E}^{(\text{inside})}(\boldsymbol{r}, t) \cong (\cos\theta / \cos\theta'')(\cos\theta''\,\hat{\boldsymbol{x}} - \sin\theta''\,\hat{\boldsymbol{z}})\exp[i(2\pi/\lambda_0)(x\sin\theta - ct)]. \tag{12}$$

The $E$ field inside the thin slab thus has a magnitude of $\cos\theta/\cos\theta''$ and is oriented at an angle $\theta''$ relative to the $x$-axis, as shown in the figure. In contrast, the $E$ field inside the semi-infinite slab has a magnitude of $\tau$ given by Eq.(7b), and is tilted away from the $x$-axis by an angle $\theta'$. Therefore, the reflection coefficient $r$ of the thin slab given by Eq.(10) must be scaled by a weight-factor in order to account for this change of the magnitude and orientation of the internal $E$ field. The weight-factor for backward radiation is given by

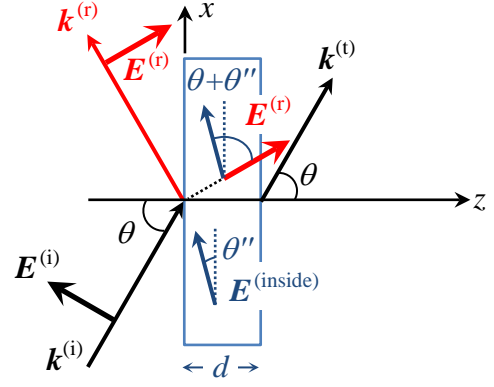

$$W_b(\theta) = \frac{\tau\cos(\theta+\theta')}{(\cos\theta/\cos\theta'')\cos(\theta+\theta'')} = \frac{\tau\cos(\theta+\theta')}{(\cos\theta/\cos\theta'')(\cos\theta\cos\theta''-\sin\theta\sin\theta'')} = \frac{\tau\cos(\theta+\theta')}{\cos\theta(\cos\theta-\sin\theta\tan\theta'')}$$

$$= \frac{\tau\cos(\theta+\theta')}{\cos\theta(\cos\theta - n^{-2}\sin\theta\tan\theta)} = \frac{\tau\cos(\theta+\theta')}{\cos^2\theta - \sin^2\theta'} = \frac{\tau\cos(\theta+\theta')}{\cos^2\theta - (\sin^2\theta + \cos^2\theta)\sin^2\theta'}$$

$$= \frac{\tau\cos(\theta+\theta')}{\cos^2\theta\cos^2\theta' - \sin^2\theta\sin^2\theta'} = \frac{\tau\cos(\theta+\theta')}{\cos(\theta-\theta')\cos(\theta+\theta')} = \frac{\tau}{\cos(\theta-\theta')}. \tag{13}$$

Similarly, the weight-factor for forward radiation by the thin slab is found to be

$$W_f(\theta) = \frac{\tau\cos(\theta-\theta')}{(\cos\theta/\cos\theta'')\cos(\theta+\theta'')} = \frac{\tau}{\cos(\theta+\theta')}. \tag{14}$$

The proof of the Ewald-Oseen theorem is now straightforward. The reflection coefficient of the semi-infinite slab is given by

$$\rho = \int_0^\infty \frac{\tau\exp(i2\pi nz\cos\theta'/\lambda_0)}{\cos(\theta-\theta')} \times \frac{i\pi(n^2\cos^2\theta - \cos^2\theta')}{\lambda_0\cos\theta} \times \exp\left(\frac{i2\pi z\cos\theta}{\lambda_0}\right) dz$$

$$= -\frac{2\cos\theta}{\cos(\theta-\theta')(\cos\theta'+n\cos\theta)} \times \frac{i\pi(n^2\cos^2\theta - \cos^2\theta')}{\lambda_0\cos\theta} \times \frac{\lambda_0}{i2\pi(n\cos\theta'+\cos\theta)}$$

$$= \frac{\cos\theta' - n\cos\theta}{\cos(\theta-\theta')(n\cos\theta'+\cos\theta)} = \frac{\cos\theta' - n\cos\theta}{(\cos\theta\cos\theta'+\sin\theta\sin\theta')(n\cos\theta'+\cos\theta)}$$

$$= \frac{\cos\theta' - n\cos\theta}{\cos\theta' + n\cos\theta}. \tag{15}$$

This is the same expression as the Fresnel reflection coefficient given by Eq.(7a). As for the $E$ field within the semi-infinite slab at a plane $z = z_0$, we write

Incident beam → $(\cos\theta\,\hat{\boldsymbol{x}} - \sin\theta\,\hat{\boldsymbol{z}})\exp\left(\frac{i2\pi z_0\cos\theta}{\lambda_0}\right)$

Forward radiation → $+(\cos\theta\,\hat{\boldsymbol{x}} - \sin\theta\,\hat{\boldsymbol{z}})\int_0^{z_0}\frac{\tau\exp(i2\pi nz\cos\theta'/\lambda_0)}{\cos(\theta+\theta')} \times \frac{i\pi(n^2\cos^2\theta - \cos^2\theta')}{\lambda_0\cos\theta} \times \exp\left[\frac{i2\pi(z_0-z)\cos\theta}{\lambda_0}\right]dz$

Backward radiation → $+(\cos\theta\,\hat{\boldsymbol{x}} + \sin\theta\,\hat{\boldsymbol{z}})\int_{z_0}^\infty\frac{\tau\exp(i2\pi nz\cos\theta'/\lambda_0)}{\cos(\theta-\theta')} \times \frac{i\pi(n^2\cos^2\theta - \cos^2\theta')}{\lambda_0\cos\theta} \times \exp\left[\frac{i2\pi(z-z_0)\cos\theta}{\lambda_0}\right]dz$

$= (\cos\theta\,\hat{\boldsymbol{x}} - \sin\theta\,\hat{\boldsymbol{z}})\exp\left(\frac{i2\pi z_0\cos\theta}{\lambda_0}\right)$

$+(\cos\theta\,\hat{\boldsymbol{x}} - \sin\theta\,\hat{\boldsymbol{z}})\frac{n\cos\theta - \cos\theta'}{\cos(\theta+\theta')(n\cos\theta' - \cos\theta)}\left[\exp\left(\frac{i2\pi nz_0\cos\theta'}{\lambda_0}\right) - \exp\left(\frac{i2\pi z_0\cos\theta}{\lambda_0}\right)\right]$

$$-(\cos\theta\,\hat{x}+\sin\theta\,\hat{z})\frac{n\cos\theta-\cos\theta'}{\cos(\theta-\theta')(n\cos\theta'+\cos\theta)}\exp\left(\frac{i2\pi n z_0\cos\theta'}{\lambda_0}\right)$$

$$=\left[(\cos\theta\,\hat{x}-\sin\theta\,\hat{z})-(\cos\theta\,\hat{x}+\sin\theta\,\hat{z})\frac{n\cos\theta-\cos\theta'}{n\cos\theta+\cos\theta'}\right]\exp\left(\frac{i2\pi n z_0\cos\theta'}{\lambda_0}\right)$$

$$=\left(\frac{2\cos\theta\cos\theta'}{\cos\theta'+n\cos\theta}\hat{x}-\frac{2n\cos\theta\sin\theta}{\cos\theta'+n\cos\theta}\hat{z}\right)\exp\left(\frac{i2\pi n z_0\cos\theta'}{\lambda_0}\right)$$

$$=(\cos\theta'\,\hat{x}-n^2\sin\theta'\,\hat{z})\tau\exp\left(\frac{i2\pi n z_0\cos\theta'}{\lambda_0}\right). \tag{16}$$

The final expression in Eq.(16) is the $E$ field amplitude inside the semi-infinite slab at $z=z_0$, its only unusual feature being the coefficient $n^2$ of $E_z$. In fact, what is being calculated here is the $E$ field in a narrow air-gap at $z=z_0$ inside the dielectric slab. The tangential component of $E$, namely, $E_x$, is the same inside the gap and in the adjacent regions of the dielectric. However, the perpendicular component, $E_z$, is not continuous; rather, it is the corresponding displacement field, $D_z$, that must remain continuous across the gap. The $n^2=\varepsilon$ factor simply converts $D_z$ in the air-gap to $E_z$ in the adjacent dielectric. This completes the proof of the Ewald-Oseen extinction theorem.

## Appendix

In the original paper, the weight-factor $W(\theta)=\cos(\theta-\theta'')/\cos(\theta+\theta'')$ was introduced based on a geometric argument and confirmed by numerical calculations. It was argued that the reflection and transmission coefficients $r_p$ and $t_p$ of the thin slab (for $p$-polarized plane-wave incident at an angle $\theta$) are related by the following equation:

$$t_p=\exp(i2\pi d\cos\theta/\lambda_0)+W(\theta)r_p. \tag{A1}$$

The angle $\theta''$ between the slab's internal $E$ field and the $x$-axis is given by $\tan\theta''=n^{-2}\tan\theta$; see Eq.(12). Here we derive Eq.(A1) in the limit when $d\ll\lambda_0$, starting with the transmission coefficient in Eq.(6d) and using Eqs.(6a), (6b) and (10). To first order in $d/\lambda_0$, the transmitted $E$ field's complex amplitude at the exit facet of the slab is approximated as follows:

$$t\exp\left(\frac{i2\pi d\cos\theta}{\lambda_0}\right)\cong n\left[E_A\left(1+\frac{i2\pi nd\cos\theta'}{\lambda_0}\right)-E_B\left(1-\frac{i2\pi nd\cos\theta'}{\lambda_0}\right)\right]$$

$$=n(E_A-E_B)+n(E_A+E_B)\left(\frac{i2\pi nd\cos\theta'}{\lambda_0}\right)$$

$$=(1-r)+\left(\frac{i2\pi n^2 d\cos\theta}{\lambda_0}\right)(1+r)$$

$$=1+\frac{i2\pi n^2 d\cos\theta}{\lambda_0}+r\left(\frac{i2\pi n^2 d\cos\theta}{\lambda_0}-1\right)$$

$$\cong 1+\frac{i2\pi n^2 d\cos\theta}{\lambda_0}+\frac{i\pi d(n^2\cos^2\theta-\cos^2\theta')}{\lambda_0\cos\theta}\left(\frac{i2\pi n^2 d\cos\theta}{\lambda_0}-1\right)$$

$$\cong\left[\exp\left(\frac{i2\pi d\cos\theta}{\lambda_0}\right)-\frac{i2\pi d\cos\theta}{\lambda_0}\right]+\frac{i2\pi n^2 d\cos\theta}{\lambda_0}-\frac{i\pi d(n^2\cos^2\theta-\cos^2\theta')}{\lambda_0\cos\theta}$$

$$=\exp\left(\frac{i2\pi d\cos\theta}{\lambda_0}\right)+\frac{i\pi d[(n^2-2)\cos^2\theta+\cos^2\theta']}{\lambda_0\cos\theta}$$

$$\cong \exp\left(\frac{i2\pi d \cos\theta}{\lambda_0}\right) + \left[\frac{(n^2-2)\cos^2\theta+\cos^2\theta'}{n^2\cos^2\theta-\cos^2\theta'}\right]r$$

$$= \exp\left(\frac{i2\pi d \cos\theta}{\lambda_0}\right) + \left[\frac{(n^2-2)\cos^2\theta+(1-n^{-2}\sin^2\theta)}{n^2\cos^2\theta-(1-n^{-2}\sin^2\theta)}\right]r$$

$$= \exp\left(\frac{i2\pi d \cos\theta}{\lambda_0}\right) + \left[\frac{(n^2-2)+(1+\tan^2\theta)-n^{-2}\tan^2\theta}{n^2-(1+\tan^2\theta)+n^{-2}\tan^2\theta}\right]r$$

$$= \exp\left(\frac{i2\pi d \cos\theta}{\lambda_0}\right) + \left[\frac{(n^2-1)+n^{-2}(n^2-1)\tan^2\theta}{(n^2-1)-n^{-2}(n^2-1)\tan^2\theta}\right]r$$

$$= \exp\left(\frac{i2\pi d \cos\theta}{\lambda_0}\right) + \left(\frac{1+n^{-2}\tan^2\theta}{1-n^{-2}\tan^2\theta}\right)r$$

$$= \exp\left(\frac{i2\pi d \cos\theta}{\lambda_0}\right) + \left(\frac{1+\tan\theta\tan\theta"}{1-\tan\theta\tan\theta"}\right)r$$

$$= \exp\left(\frac{i2\pi d \cos\theta}{\lambda_0}\right) + \left(\frac{\cos\theta\cos\theta"+\sin\theta\sin\theta"}{\cos\theta\cos\theta"-\sin\theta\sin\theta"}\right)r$$

$$= \exp\left(\frac{i2\pi d \cos\theta}{\lambda_0}\right) + \frac{\cos(\theta-\theta")}{\cos(\theta+\theta")}r. \tag{A2}$$

The last line of Eq.(A2) is the same as Eq.(A1). The proof is thus complete.